\documentclass[12pt]{article}
\usepackage{times}
\usepackage{epsfig}
\usepackage{amsmath}
\usepackage{amsfonts}

\def\beq{\begin{equation}}
\def\eeq{\end{equation}}
\def\bea{\begin{eqnarray}}
\def\eea{\end{eqnarray}}

\def\nn{\nonumber}
\def\Eq#1{Eq.~(\ref{#1})}

\def\pb{p\hspace{-.42em}/\hspace{-.07em}}
\def\qb{q\hspace{-.42em}/\hspace{-.07em}}
\def\td#1{\tilde{\delta}\left(#1\right)}

\newcommand{\la}{\langle}
\newcommand{\ra}{\rangle}
\def\bom#1{{\mbox{\boldmath $#1$}}}
\def\sp{{\bom {Sp}}}
\def\ket#1{|{#1}\ra}
\def\bra#1{\la{#1}|}
\def\qb{\mathbf{q}}
\def\pb{\mathbf{p}}

\begin{document}

\begin{titlepage}
\renewcommand{\thefootnote}{\fnsymbol{footnote}}
\begin{flushright}
     LPN14-078  \\
     IFIC/13-85
     \end{flushright}
\par \vspace{10mm}
\begin{center}
{\LARGE \bf
On the singular behaviour of scattering amplitudes in quantum 
field theory}
\end{center}
\par \vspace{2mm}
\begin{center}
{\bf 
Sebastian Buchta~$^{(a)}$\footnote{E-mail: sbuchta@ific.uv.es}, 
Grigorios Chachamis~$^{(a)}$\footnote{E-mail: grigorios.chachamis@ific.uv.es}, 
Petros Draggiotis~$^{(b)}$\footnote{E-mail: petros.draggiotis@gmail.com}, \\
Ioannis Malamos~$^{(a)}$\footnote{E-mail: ioannis.malamos@ific.uv.es}, 
and
Germ\'an Rodrigo~$^{(a)}$\footnote{E-mail: german.rodrigo@csic.es}
}
\vspace{5mm}

${}^{(a)}$ Instituto de F\'{\i}sica Corpuscular, 
Universitat de Val\`{e}ncia --  
Consejo Superior de Investigaciones Cient\'{\i}ficas, 
Parc Cient\'{\i}fic, E-46980 Paterna, Valencia, Spain \\
\vspace*{2mm}
${}^{(b)}$ Institute of Nuclear and Particle Physics, NCSR "Demokritos", 
Agia Paraskevi, 15310, Greece \\
\vspace*{2mm}
\end{center}

\par \vspace{2mm}
\begin{center} {\large \bf Abstract} \end{center}
\begin{quote}
We analyse the singular behaviour of one-loop integrals and 
scattering amplitudes in the framework of the loop--tree duality approach. 
We show that there is a partial cancellation of singularities 
at the loop integrand level among the different
components of the corresponding dual representation
that can be interpreted in terms of causality. 
The remaining threshold and infrared singularities are restricted 
to a finite region of the loop momentum space, which is of the 
size of the external momenta and can be mapped to
the phase-space of real corrections to cancel 
the soft and collinear divergences.  
\end{quote}

\par \vspace{5mm}

\vspace*{\fill}
\begin{flushleft}
October 21, 2014
\end{flushleft}
\end{titlepage}

\setcounter{footnote}{0}
\renewcommand{\thefootnote}{\fnsymbol{footnote}}

\section{Introduction}

The recent discovery of the Higgs boson at the LHC represents
a great success of the Standard Model (SM) of elementary particles. 
While at the same time, the absence so far of a clear signal 
of physics beyond the SM leaves a certain degree of dissatisfaction.
These two facts, together with the high quality of data that the LHC
will provide in the next run, increases the relevance of high-precision 
theoretical predictions for the analysis of known phenomena and for finding 
innovative strategies to achieve new discoveries. 

The domain of perturbative calculations in quantum field theories, e.g.
the SM and beyond, has shown an extraordinary progress in the recent years. 
Today, $2\to 4$ processes at next-to-leading order (NLO) are state of the 
art~\cite{Berger:2009zg,Melnikov:2009wh,Bevilacqua:2010ve,Denner:2010jp,Campanario:2013qba},
and even higher multiplicities are affordable~\cite{Bern:2013gka}. 
Several tools for the automated calculation of NLO differential 
cross sections are available~\cite{Bevilacqua:2011xh,Cullen:2014yla},
including the merging with parton showers~\cite{Alwall:2014hca}.
There has been also a lot of advances in next-to-next-to-leading order (NNLO) 
calculations~\cite{Bolzoni:2010xr,Catani:2010en,Catani:2009sm,Anastasiou:2007mz,Czakon:2013goa}.
Still, besides ultraviolet singularities which are easily removed by renormalization,
the cancellation of infrared singularities by 
the coherent sum over different real and virtual soft and 
collinear partonic configurations in the final state 
is at the core and the main source of cumbersomeness
of any perturbative calculation at higher 
orders~\cite{Catani:1996vz,Catani:1996jh,Frixione:1995ms,GehrmannDeRidder:2005cm,Catani:2007vq}.  

The loop--tree duality method~\cite{Catani:2008xa,Bierenbaum:2010cy,Bierenbaum:2012th,Bierenbaum:2013nja} 
establishes that generic loop quantities (loop integrals and scattering amplitudes) 
in any relativistic, local and unitary field theory can be written 
as a sum of tree-level objects obtained after making all possible cuts to 
the internal lines of the corresponding Feynman diagrams, with one single cut per loop
and integrated over a measure that closely 
resembles the phase-space of the corresponding real corrections. 
This duality relation is realized by a modification of 
the customary +i0 prescription of the Feynman propagators.
At one-loop, the new prescription compensates for the absence of multiple-cut 
contributions that appear in the Feynman Tree Theorem~\cite{Feynman:1963ax,F2}.
The modified phase-space raises the intriguing possibility
that virtual and real corrections can be brought together under a 
common integral and treated with Monte Carlo techniques at the same time.
In this paper we analyse the singular behaviour of one-loop integrals and 
scattering amplitudes in the framework of the loop--tree duality method. 
On the one hand, working in the loop momentum space is an attractive approach 
because it allows a rather direct physical interpretation of the singularities 
of the loop quantities~\cite{Sterman:1978bi}. On the other hand, the 
possibility to relate virtual and real corrections opens an interesting 
line to understand explicitly the cancellation of infrared singularities. 

The outline of the paper is as follows. 
In Section~\ref{sec:loop} we discuss the singular behaviour of 
scalar loop integrals in the loop momentum space. 
In Section~\ref{sec:cancel} we prove that there is a partial 
cancellation of singularities at the integrand level 
among different contributions of the dual representation 
of a loop integral. In Section~\ref{sec:real}, 
collinear factorization is used to sketch a phase-space mapping 
between virtual and real corrections for the local cancellation 
of infrared divergences. Finally, conclusions and outlook are 
presented in Section~\ref{sec:conclusions}.

\section{The singular behaviour of the loop integrand}
\label{sec:loop}

We consider a general one-loop $N$-leg scalar integral
\beq
\label{Ln}
L^{(1)}(p_1, p_2, \dots, p_N) =
\int_{\ell} \, \prod_{i \in \alpha_1} \, G_F(q_i)~, \qquad
\int_{\ell} \bullet =-i \int \frac{d^d \ell}{(2\pi)^{d}} \; \bullet~,
\eeq
where 
\beq
G_F(q_i)=\frac{1}{q_i^2-m_i^2+i0}
\label{eq:feynman}
\eeq
are Feynman propagators that depend on the 
loop momentum $\ell$, which flows anti-clockwise, 
and the four-momenta of the external legs $p_{i}$, 
$i \in \alpha_1 = \{1,2,\ldots N\}$, which are taken as outgoing and 
are ordered clockwise. 
We use dimensional regularization with $d$  
the number of space-time dimensions. 
The momenta of the internal lines $q_{i,\mu} = (q_{i,0},\mathbf{q}_i)$, 
where $q_{i,0}$ is the energy (time component) and $\qb_{i}$ are 
the spacial components, are defined as $q_{i} = \ell + k_i$ with 
$k_{i} = p_{1} + \ldots + p_{i}$, and $k_{N} = 0$ by momentum conservation. 
We also define $k_{ji} = q_j - q_i$.

The loop integrand becomes singular in regions of the 
loop momentum space in which subsets of internal lines go on-shell,
although the existence of singular points of the integrand 
is not enough to ensure the emergence in the loop integral
of divergences in the dimensional regularization parameter. 
Nevertheless, numerical integration over integrable singularities 
still requires a contour deformation~\cite{Gong:2008ww,Nagy:2006xy,Kramer:2002cd,Soper:2001hu,Soper:1999xk,Soper:1998ye,Becker:2012nk,Becker:2012bi}, 
namely, to promote the loop momentum to the complex 
plane in order to smoothen the loop matrix elements in the singular 
regions of the loop integrand. Hence, the relevance to 
identify accurately all the integrand singularities. 

\begin{figure}[ht]
\begin{center}
\includegraphics[width=7cm]{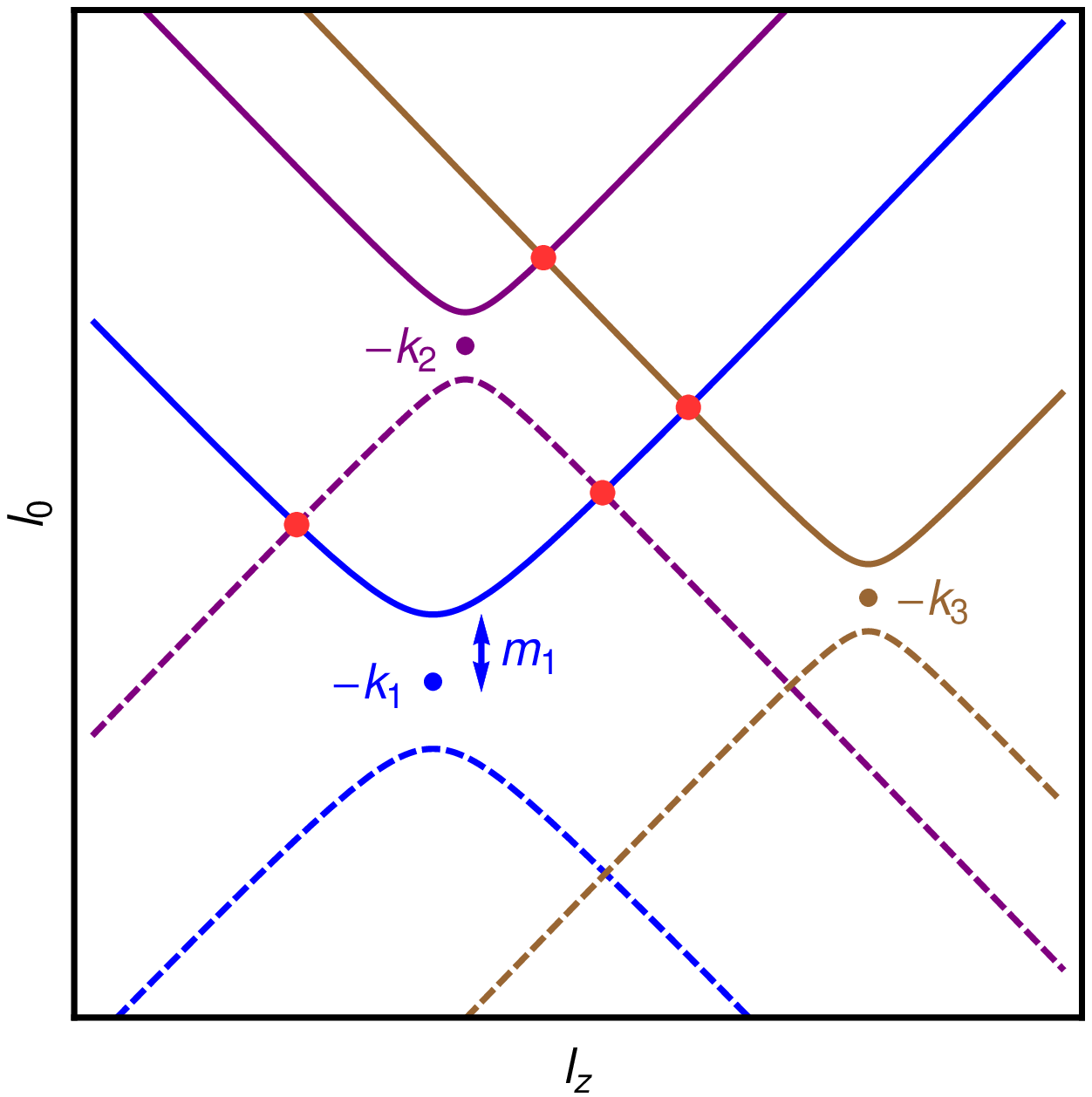}
\includegraphics[width=7cm]{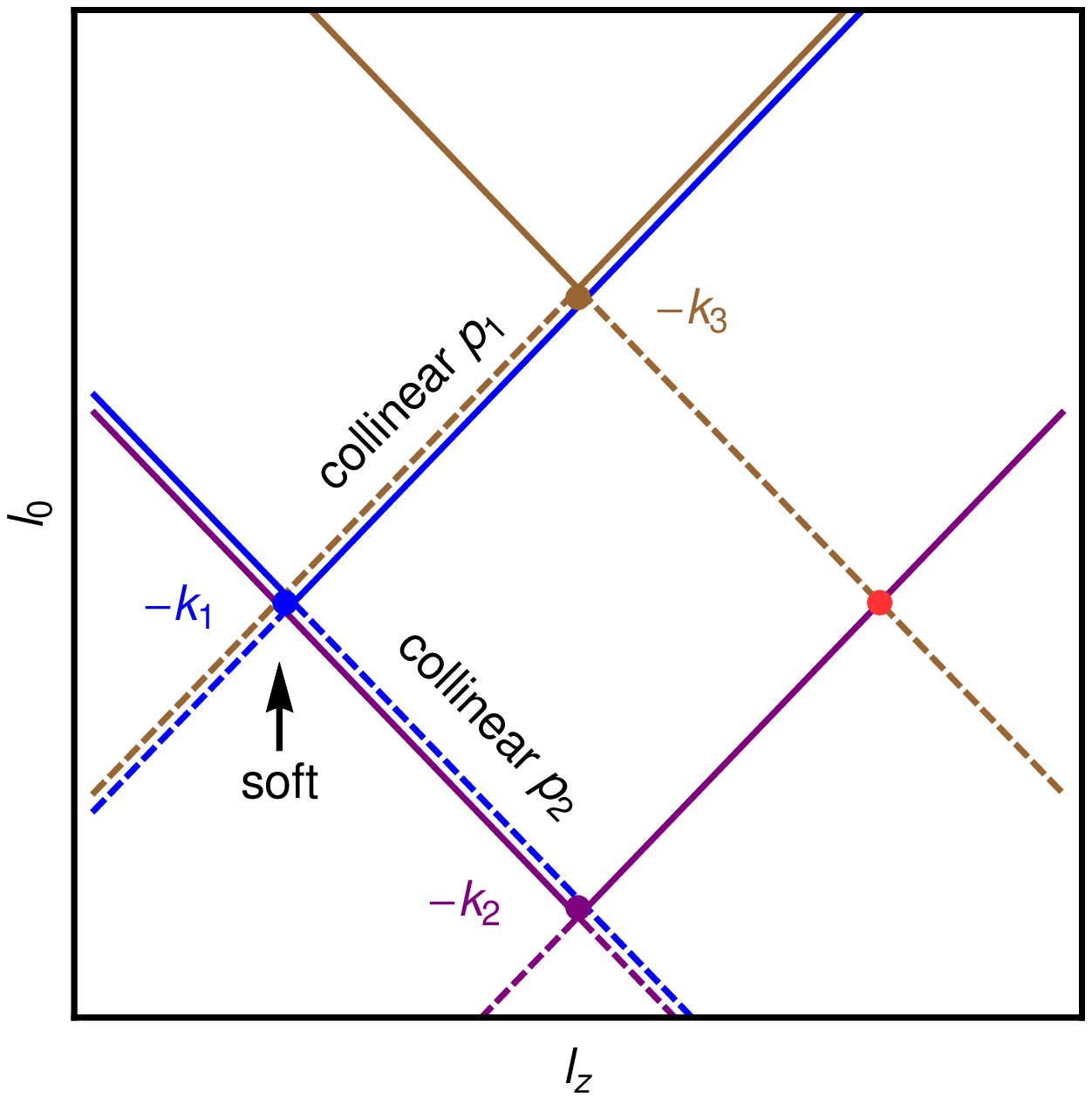}
\caption{On-shell hyperboloids for three arbitrary propagators 
in Cartesian coordinates in the ($\ell_0$,$\ell_z$) space (left). 
Kinematical configuration with infrared singularities (right).
In the latter case, the on-shell hyperboloids degenerate to light-cones. 
\label{fig:cartesean}}
\end{center}
\end{figure}

In Cartesian coordinates, the Feynman propagator in~\Eq{eq:feynman}  
becomes singular at hyperboloids with origin in $-k_{i}$, 
where the minimal distance between each hyperboloid and 
its origin is determined by the internal mass $m_i$.
This is illustrated in Fig.~\ref{fig:cartesean}, where for simplicity
we work in $d=2$ space-time dimensions. Figure~\ref{fig:cartesean}~(left)
shows a typical kinematical situation where two 
momenta, $k_1$ and $k_2$, are separated by a time-like distance, 
$k_{21}^2 > 0$, and a third momentum, $k_3$, is space-like separated  
with respect to the other two, 
$k_{31}^2 <0$ and $k_{32}^2 <0$. The on-shell forward hyperboloids
($q_{i,0}>0$) are represented in Fig.~\ref{fig:cartesean} by solid lines, 
and the backward hyperboloids ($q_{i,0}<0$) by dashed lines. 
For the discussion that will follow it is important to stress that Feynman 
propagators become positive inside the respective 
hyperboloid and negative outside. 
Two or more Feynman propagators become simultaneously singular 
where their respective hyperboloids intersect. 
In most cases, these singularities, due to normal or anomalous 
thresholds~\cite{Mandelstam:1960zz,Rechenberg:1972rq}
of intermediate states, are integrable.
However, if two massless propagators are separated by a 
light-like distance, $k_{ji}^2=0$, then the overlap 
of the respective light-cones is tangential, 
as illustrated in Fig.~\ref{fig:cartesean}~(right),
and leads to non-integrable collinear singularities. 
In addition, massless propagators can generate soft singularities 
at $q_{i}=0$. 

The dual representation of the scalar one-loop integral in \Eq{Ln}
is the sum of $N$ dual integrals~\cite{Catani:2008xa,Bierenbaum:2010cy}:
\bea
\label{oneloopduality}
L^{(1)}(p_1, p_2, \dots, p_N) 
&=& - \sum_{i \in \alpha_1} \, \int_{\ell} \; \td{q_i} \,
\prod_{\substack{j \in \alpha_1 \\ j\neq i}} \,G_D(q_i;q_j)~,
\eea 
where
\beq
G_D(q_i;q_j) = \frac{1}{q_j^2 -m_j^2 - i0 \, \eta \, k_{ji}}
\eeq
are the so-called dual propagators, as defined in Ref.~\cite{Catani:2008xa},
with $\eta$ a {\em future-like} vector, $\eta^2 \ge 0$, 
with positive definite energy $\eta_0 > 0$.
The delta function 
$\td{q_i} \equiv 2 \pi \, i \, \theta(q_{i,0}) \, \delta(q_i^2-m_i^2)$
sets the internal lines on-shell by selecting the pole of the propagators
with positive energy $q_{i,0}$ and negative imaginary part. 
In the following we take $\eta_\mu = (1,\mathbf{0})$, and thus 
$- i0 \, \eta \, k_{ji} = -i0 \, k_{ji,0}$. This is equivalent 
to performing the loop integration along the on-shell forward hyperboloids.
Let us mention that in the light-cone coordinates 
($\ell_+$, $\ell_-$, $\mathbf{l}_\perp$), 
where $\ell_\pm=(\ell_0 \pm \ell_{d-1})/\sqrt{2}$,
Feynman propagators vanish at hyperboloids in the plane ($\ell_+$,$\ell_-$) 
which are similar to those depicted in Fig.~\ref{fig:cartesean} but 
rotated by 45 degrees. Consequently, by selecting the forward hyperboloids 
the integration limits of either $\ell_+$ or $\ell_-$ are restricted 
and the restrictions are different for each dual integral.
For this reason, although~\Eq{oneloopduality}
is valid for any system of coordinates, we will stick 
for the rest of the paper to Cartesian coordinates where 
all the dual integrals share the same integration limits
for the loop three-momentum. 

A crucial point of our discussion is the observation that 
dual propagators can be rewritten as 
\beq
\td{q_i} \, G_D(q_i;q_j) = 
i\, 2 \pi \, 
\frac{\delta(q_{i,0}-q_{i,0}^{(+)})}{2 q_{i,0}^{(+)}} \, 
\frac{1}{(q_{i,0}^{(+)} + k_{ji,0})^2-(q_{j,0}^{(+)})^2}~,
\label{eq:newdual}
\eeq
where
\beq
q_{i,0}^{(+)} = \sqrt{\mathbf{q}_i^2 + m_i^2-i0}
\label{qi0distance}
\eeq
is the loop energy measured along the on-shell hyperboloid 
with origin at $-k_i$. By definition we have ${\rm Re}(q_{i,0}^{(+)}) \ge 0$. 
The factor $1/q_{i,0}^{(+)}$ can become singular for $m_i=0$, but the integral 
$\int_\ell \delta(q_{i,0}-q_{i,0}^{(+)})/q_{i,0}^{(+)}$ is still convergent by two 
powers in the infrared. 
Soft singularities require two dual propagators, where each of the two
dual propagators contributes with one power in the infrared.
From \Eq{eq:newdual} it is obvious that dual propagators become 
singular, $G_D^{-1}(q_i;q_j)=0$, if one of the following conditions is fulfilled:
\bea
&& q_{i,0}^{(+)}+q_{j,0}^{(+)}+k_{ji,0}=0~, \label{ellipsoid} \\
&& q_{i,0}^{(+)}-q_{j,0}^{(+)}+k_{ji,0}=0~. \label{hyperboloid}
\eea
The first condition, \Eq{ellipsoid}, is satisfied if the 
forward hyperboloid of $-k_i$ intersects 
with the backward hyperboloid of $-k_j$. 
The second condition, \Eq{hyperboloid}, is true when 
the two forward hyperboloids intersect each other. 

In the massless case, \Eq{ellipsoid} and \Eq{hyperboloid} are the equations
of conic sections in the loop three-momentum space; 
$q_{i,0}^{(+)}$ and $q_{j,0}^{(+)}$ are the distance to the {\it foci} 
located at $-\mathbf{k}_i$ and $-\mathbf{k}_j$, respectively, 
and the distance between the foci is $\sqrt{\mathbf{k}_{ji}^2}$.
If internal masses are non-vanishing, \Eq{qi0distance} can be reinterpreted 
as the distance associated to a four-dimensional space with 
one ``massive'' dimension and the foci now located 
at $(-\mathbf{k}_i,-m_i)$ and $(-\mathbf{k}_j,-m_j)$, respectively. 
Then, the singularity arises at the intersection of 
the conic sections given by \Eq{ellipsoid} or \Eq{hyperboloid} 
in this generalized space with the zero mass plane. This picture 
is useful to identify the singular regions of the loop integrand 
in the loop three-momentum space.

The solution to \Eq{ellipsoid} is an ellipsoid and clearly requires $k_{ji,0}<0$.
Moreover, since it is the result of the intersection of a forward 
with a backward hyperboloid the distance between the two 
propagators has to be future-like, $k_{ji}^2 \ge 0$. 
Actually, internal masses restrict this condition. Bearing in mind 
the image of the conic sections in the generalized massive space so 
we can deduce intuitively that \Eq{ellipsoid} has solution for 
\beq
k_{ji}^2-(m_j+m_i)^2 \ge 0~, \qquad k_{ji,0}<0~, \qquad 
\rm{forward~with~backward~hyperboloids}~.
\label{eq:generalizedtimelike}
\eeq
The second equation, \Eq{hyperboloid}, leads to a hyperboloid in the 
generalized space, and there are solutions for $k_{ji,0}$ either 
positive or negative, namely when either of the two momenta are 
set on-shell. However, by interpreting the result in the generalized 
space it is clear that the intersection with the zero mass plane 
does not always exist, and if it exists, it can be either an ellipsoid 
or a hyperboloid in the loop three-momentum space. 
Here, the distance between the momenta of the propagators 
has to be space-like, although also time-like configurations 
can fulfil~\Eq{hyperboloid} as far as the time-like 
distance is small or close to light-like. 
The following condition is necessary:
\beq
k_{ji}^2-(m_j-m_i)^2 \le 0~, \qquad \rm{two~forward~hyperboloids}~.
\label{eq:generalizedspacelike}
\eeq
In any other configuration, the singularity appears for loop 
three-momenta with imaginary components. 

\section{Cancellation of singularities among dual integrands}
\label{sec:cancel}

In this section we prove one of the main properties of the loop--tree 
duality method, namely the partial cancellation of 
singularities among different dual integrands.
This represents a significant advantage with respect to the integration 
of regular loop integrals in the $d$-dimensional space, where one 
single integrand cannot obviously lead to such cancellation. 

Let's consider first two Feynman propagators separated by 
a space-like distance, $k_{ji}^2 < 0$ (or more generally 
fulfilling~\Eq{eq:generalizedspacelike}). 
In the corresponding dual representation one of these
propagators is set on-shell and the other becomes dual, 
and the integration occurs along the respective on-shell forward 
hyperboloids. See again Fig.~\ref{fig:cartesean}~(left) for a graphical 
representation of this set-up.
There, the two forward hyperboloids of $-k_1$ and $-k_3$ intersect 
at a single point. Integrating over $\ell_z$ along the forward 
hyperboloid of $-k_1$ we find that the dual propagator 
$G_D(q_1;q_3)$, which is negative below the intersection point where 
the integrand becomes singular, changes sign above this point 
as we move from outside to inside the on-shell hyperboloid of $-k_3$. 
The opposite occurs if we set $q_3$ on-shell;
$G_D(q_3;q_1)$ is positive below the intersection point, and 
negative above. The change of sign leads to the cancellation of the 
common singularity. Notice that also the dual $i0$ prescription 
changes sign. In order to prove 
analytically this cancellation, we define 
$x = q_{i,0}^{(+)} - q_{j,0}^{(+)} + k_{ji,0}$. 
In the limit $x\to 0$:
\beq
\lim_{x \to 0} \, \left( 
\td{q_i} \, G_D(q_i;q_j) + (i \leftrightarrow j) 
\right)
= \left( \frac{1}{x}-\frac{1}{x} \right) \, 
\frac{1}{2 q_{j,0}^{(+)}} \, \td{q_i} + {\cal O}(x^0)~,
\label{eq:spacecancel1}
\eeq
and thus the leading singular behaviour cancels
among the two dual contributions. 
The cancellation of these singularities is not altered by 
the presence of other non-vanishing dual propagators 
(neither by numerators) because
\beq
\lim_{x \to 0} \, G_D(q_j;q_k) = 
\lim_{x \to 0} \, 
\frac{1}{(q_{j,0}^{(+)} + k_{ki,0} - k_{ji,0})^2-(q_{k,0}^{(+)})^2} =
\lim_{x \to 0} \, G_D(q_i;q_k)~,
\label{eq:spacecancel2}
\eeq
where we have used the identity $k_{kj,0} = k_{ki,0} - k_{ji,0}$. 
If instead, the separation is time-like (in the sense 
of~\Eq{eq:generalizedtimelike}), 
we define $x = q_{i,0}^{(+)} + q_{j,0}^{(+)} + k_{ji,0}$, and find
\beq
\lim_{x \to 0} \, \left( 
\td{q_i} \, G_D(q_i;q_j) + (i \leftrightarrow j) 
\right)
= - \theta(-k_{ji,0}) \, 
\frac{1}{x} \, \frac{1}{2 q_{j,0}^{(+)}} \, 
\td{q_i} + (i \leftrightarrow j) + {\cal O}(x^0)~.
\label{eq:timecancel}
\eeq
In this case the singularity of the integrand remains because of the 
Heaviside step function.

We should consider also the case in which more than two 
propagators become simultaneously singular. 
To analyse the intersection of three forward 
hyperboloids, we define 
\beq
\lambda \, x = q_{i,0}^{(+)} - q_{j,0}^{(+)} + k_{ji,0}~, \qquad
\lambda \, y = q_{i,0}^{(+)} - q_{k,0}^{(+)} + k_{ki,0}~.
\eeq
As before, we use the identity $k_{kj,0} = k_{ki,0} - k_{ji,0}$, 
and thus $q_{j,0}^{(+)} - q_{k,0}^{(+)} + k_{kj,0} = \lambda \, (y-x)$.
In the limit in which the three propagators become simultaneously singular:
\bea
&& \lim_{\lambda \to 0}
\left( \td{q_i} \, G_D(q_i;q_j) \, G_D(q_i;q_k) + {\rm perm.} \right) 
= \nn \\ &&
\frac{1}{\lambda^2} \,
\left( \frac{1}{x\, y} + \frac{1}{x\, (x-y)} + \frac{1}{y\, (y-x)} \right) \, 
\frac{1}{2 q_{j,0}^{(+)}} \, \frac{1}{2 q_{k,0}^{(+)}} \, 
\td{q_i} + {\cal O}(\lambda^{-1})~,
\label{eq:threespacecancel}
\eea
and again the leading singular behaviour cancels in the sum. 
Although not shown for simplicity in \Eq{eq:threespacecancel}, 
also the ${\cal O}(\lambda^{-1})$ terms cancel in the sum, thus 
rendering the integrand finite in the limit $\lambda \to 0$.
For three propagators there are also more possibilities:
two forward hyperboloids might intersect simultaneously with a backward 
hyperboloid, or two backward hyperboloids might intersect with a forward 
hyperboloid. In the former case, we define 
$\lambda \, x = q_{i,0}^{(+)} + q_{k,0}^{(+)} + k_{ki,0}$, and
$\lambda \, y = q_{j,0}^{(+)} + q_{k,0}^{(+)} + k_{kj,0}$,
with $k_{ki,0}<0$ and $k_{kj,0}<0$, and hence 
$q_{i,0}^{(+)} - q_{j,0}^{(+)} + k_{ji,0} = \lambda (x-y)$. 
In the $\lambda \to 0$ limit 
\bea
&& \lim_{\lambda \to 0}
\left( \td{q_i} \, G_D(q_i;q_j) \, G_D(q_i;q_k) + {\rm perm.}
\right) = \nn \\ &&
\theta(-k_{ki,0}) \, \theta(-k_{kj,0}) \, 
\frac{1}{\lambda^2} \, \left( \frac{1}{x\, (y-x)} + \frac{1}{y\, (x-y)} \right) \, 
\frac{1}{2 q_{j,0}^{(+)}} \, \frac{1}{2 q_{k,0}^{(+)}} \, 
\td{q_i} + {\cal O}(\lambda^{-1})~.
\label{eq:spacecancel3}
\eea
Notice that the singularity in $1/(x-y)$ cancels in \Eq{eq:spacecancel3} 
(also at ${\cal O}(\lambda^{-1})$). 
In the latter case, we set as before $\lambda \, x = q_{i,0}^{(+)} + q_{k,0}^{(+)} + k_{ki,0}$, 
and define $\lambda \, z = q_{i,0}^{(+)} + q_{j,0}^{(+)} + k_{ji,0}$, then 
\bea
&& \lim_{\lambda \to 0}
\left( \td{q_i} \, G_D(q_i;q_j) \, G_D(q_i;q_k) + {\rm perm.}
\right) = 
- \theta(-k_{ki,0}) \, \nn \\ && \times \theta(-k_{ji,0}) \, 
\frac{1}{\lambda^2} \, \left( \frac{1}{x\, z} \right) \,
\frac{1}{2 q_{j,0}^{(+)}} \, \frac{1}{2 q_{k,0}^{(+)}} \, 
\td{q_i} + {\cal O}(\lambda^{-1})~. 
\label{eq:spacecancel3b}
\eea
Similarly, it is straightforward to 
prove that four forward hyperboloids
do not lead to any common singularity and more generally  
that the remaining multiple singularities are only driven 
by propagators that are time-like connected and less energetic 
than the propagator which is set on-shell. 

Thus, we conclude that singularities 
of space-like separated propagators~\footnote{
Including light-like and time-like configurations such that 
\Eq{eq:generalizedspacelike} is fulfilled.}, 
occurring in the intersection of on-shell forward hyperboloids,
are absent in the dual representation of the loop integrand. 
The cancellation of these singularities 
at the integrand level already represents a big advantage of 
the loop--tree duality with respect to the direct integration in the 
four dimensional loop space; it makes unnecessary the use of contour 
deformation to deal numerically with the integrable singularities of 
these configurations. 
This conclusion is also valid for loop scattering amplitudes. 
Moreover, this property can be extended in a straightforward manner 
to prove the partial cancellation of infrared singularities.

Collinear singularities occur when two massless propagators are 
separated by a light-like distance, $k_{ji}^2=0$. In that case, the 
corresponding light-cones overlap tangentially along 
an infinite interval. Assuming $k_{i,0} > k_{j,0}$, however,  
the collinear singularity for $\ell_0 > -k_{j,0}$ appears 
at the intersection of the two forward light-cones, 
with the forward light-cone of $-k_{j}$ located
inside the forward light-cone of $-k_{i}$, 
or equivalently, with the forward light-cone 
of $-k_{i}$ located outside the forward light-cone of $-k_{j}$, 
Thus, the singular behaviour of the two dual components 
cancel against each other, following the same qualitative 
arguments given before. 
For $-k_{i,0} < \ell_0 < -k_{j,0}$, instead, it is the forward 
light-cone of $-k_i$ that intersects tangentially 
with the backward light-cone of $-k_j$ according to~\Eq{ellipsoid}. 
The collinear divergences survive in this energy strip, 
which indeed also limits the range of the loop three-momentum 
where infrared divergences can arise. 
If there are several reference momenta separated by light-like distances
the infrared strip is limited by the minimal and maximal energies
of the external momenta. The soft singularity of the integrand
at $q_{i,0}^{(+)}=0$ leads to soft divergences only if two other 
propagators, each one contributing with one power in the infrared, 
are light-like separated from $-k_i$. In Fig.~\ref{fig:cartesean}~(right)
this condition is fulfilled only at $q_{1,0}^{(+)}=0$, 
but not at $q_{2,0}^{(+)}=0$ neither at $q_{3,0}^{(+)}=0$.

In summary, both threshold and infrared singularities are 
constrained in the dual representation of the loop integrand 
to a finite region where the loop three-momentum is of the order 
of the external momenta. Singularities outside this region, 
occurring in the intersection of on-shell forward hyperboloids
or light-cones, cancel in the sum of all the dual contributions. 

\section{Cancellation of infrared singularities with real corrections}
\label{sec:real}

Having constrained the loop singularities to a finite region of the 
loop momentum space, we discuss now how to map this region into the 
finite-size phase-space of the real corrections for the cancellation 
of the remaining infrared singularities.
The use of collinear factorization and splitting matrices, 
encoding the collinear singular behaviour of scattering amplitudes
as introduced in Ref.~\cite{Catani:2003vu,Sborlini:2013jba}, 
is suitable for this discussion.

\begin{figure}[thb]
\begin{center}
\includegraphics[width=15cm]{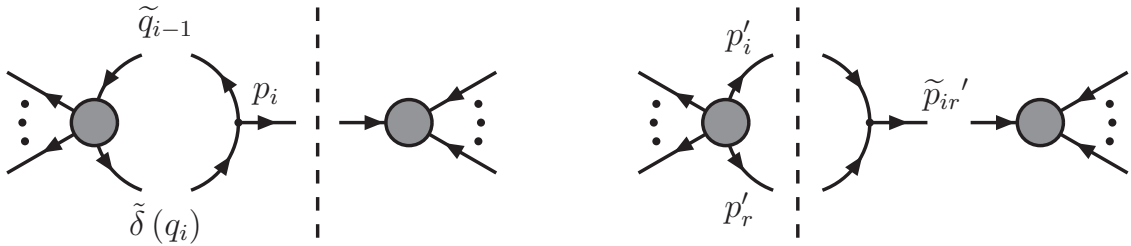}
\caption{\label{fig:collinear} 
{\em Factorization of the dual one-loop and tree-level squared amplitudes 
in the collinear limit. The dashed line represents the momentum conservation
cut.}}
\end{center}
\end{figure}

We consider the interference of the one-loop scattering amplitude
${\cal M}^{(1)}_N$ with the corresponding $N$-parton 
tree-level scattering amplitude ${\cal M}^{(0)}_N$, 
which is integrated with the appropriate phase-space factor 
\beq
\int d\Phi_N(p_1; p_2,\ldots, p_N) = 
\left(\prod_{i=2}^N \int_{p_i} \td{p_i} \right) \, 
(2\pi)^d \, \delta^{(d)} (\sum_{i=1}^{N} p_i)~,
\eeq
where we assume that only the external momentum $p_1$ is incoming ($p_{1,0}<0$).
Then, we select the corresponding dual contribution with the internal massless 
line $q_i$ on-shell
\bea
I^{(1)}_{i} &=& 
2 {\rm Re} \, \int d\Phi_N(p_1; p_2,\ldots, p_N) 
\int_{\ell}  \,  \td{q_i} \, \theta(p_{i,0}-q_{i,0}^{(+)}) \nn \\ &\times&  
\bra{{\cal M}^{(0)}_N (p_1, \ldots, p_{N})} \, 
{\cal M}^{(0)}_{N+2} (\ldots, p_i, -q_i, q_i, p_{i+1}, \ldots) \ra~, 
\label{eq:I1ith}
\eea
where the loop energy in~\Eq{eq:I1ith} is restricted 
by the energy of the adjacent external massless particle $p_{i,0}$
to select the infrared sector, according to the discussion of the previous sections.
We also consider the $N+1$-parton tree-level scattering amplitude 
\beq
\ket{{\cal M}^{(0),\, ir}_{N+1}(p_1, p_2', \ldots)} = 
\ket{{\cal M}^{(0),\, ir}_{N+1} (\ldots, p_{ir}' \to p_i'+p_r', \ldots)}~, 
\label{eq:trees}
\eeq
where an extra particle is radiated from parton $i$, 
with $p_{ir}' = p_i' + p_r'$, and the complementary 
scattering amplitude ${\cal M}^{(0)}_{N+1}$
that contains all the tree-level contributions with the exception 
of those already included in ${\cal M}^{(0),\, ir}_{N+1}$.
The corresponding interference, integrated over the 
phase-space of the final-state particles, is 
\bea
&& I^{(0)}_{ir} = 2 {\rm Re} \, 
\int d\Phi_{N+1}(p_1; p_2', \ldots) \,
\bra{{\cal M}^{(0),\, ir}_{N+1} (p_1, p_2', \ldots)} \, 
{\cal M}^{(0)}_{N+1} (p_1, p_2', \ldots) \ra~.
\label{eq:I0ab}
\eea
For the simplicity of the presentation, 
we do not consider explicitly in this paper 
the square of ${\cal M}^{(0),\, ir}_{N+1}$, which is related with 
a self-energy insertion in an external leg and whose infrared 
divergences are removed by wave-function 
remormalization~\cite{Catani:2008xa}.  
The final-state external momenta of the loop 
and tree amplitudes in \Eq{eq:I1ith} and \Eq{eq:I0ab}, 
although labelled with the same indices, are constrained by  
different phase-space momentum conservation delta functions. 
A mapping between the primed (real amplitudes) and unprimed 
(virtual amplitudes) momenta is necessary to show the 
cancellation of collinear divergences. 

In the limit where $\pb_i$ and $\qb_i$ become collinear the
dual one-loop matrix element ${\cal M}^{(0)}_{N+2}$ 
in \Eq{eq:I1ith} factorizes as 
\beq
\ket{{\cal M}^{(0)}_{N+2} (\ldots, p_i, -q_i, q_i, \ldots)} =
\sp^{(0)}(p_i, -q_i; -\widetilde{q}_{i-1}) \, 
\ket{\overline{\cal M}^{(0)}_{N+1} 
(\ldots, -\widetilde{q}_{i-1}, q_i, \ldots)} 
+ {\cal O}(\sqrt{q_{i-1}^2})~,
\label{eq:piqicoll}
\eeq
where the reduced matrix element $\overline{\cal M}^{(0)}_{N+1}$
is obtained by replacing the two collinear partons of 
${\cal M}^{(0)}_{N+2}$ by a single parent parton with light-like 
momentum 
\beq
\widetilde{q}_{i-1}^\mu = q_{i-1}^\mu - \frac{q_{i-1}^2 \, n^\mu}{2  \, n q_{i-1}}~, 
\eeq
with $n^\mu$ a light-like vector, $n^2=0$. 
Similarly, in the limit where $\pb_i'$ and $\pb_r'$
become collinear the tree-level matrix element 
${\cal M}^{(0),\, ir}_{N+1}$ factorizes as 
\beq
\bra{{\cal M}^{(0),\, ir}_{N+1} (p_1, p_2', \ldots, p_{N+1}')} =
\bra{\overline{\cal M}^{(0)}_N 
(\ldots, p_{i-1}', \widetilde{p}_{ir}', p_{i+1}', \ldots)} \, 
\sp^{(0) \dagger}(p_i', p_r'; \widetilde{p}_{ir}') 
+ {\cal O}(\sqrt{s_{ir}'})~,
\label{eq:piprimecoll}
\eeq
where $s_{ir}' = p_{ir}'^2$, and
\beq
\widetilde{p}_{ir}'^{\mu} = p_{ir}'^{\mu} 
- \frac{s_{ir}' \, n^\mu}{2\, n  p_{ir}'}
\eeq
is the light-like momentum of the parent parton. 
A graphical representation of the collinear limit of both virtual 
and real corrections is illustrated in Fig.~\ref{fig:collinear}.  
This graph suggests that in the collinear limit the mapping 
between the four-momenta of the virtual and real matrix elements 
should be such that 
$p_i = \widetilde{p}_{ir}'$,
$p_{j} = p_{j}' (j \ne i)$,
$-\widetilde{q}_{i-1} = p_i'$ and 
$q_i = p_r'$. 
Notice that $p_r'$ is restricted by momentum conservation 
but $q_i$ is not. However, the relevant infrared region is 
bound by $q_{i,0}^{(+)} \le p_{i,0}$  in \Eq{eq:I1ith}. This restriction 
allows to map $q_i$ to $p_r'$. The mapping, nevertheless, 
is not as obvious as can be induced from Fig.~\ref{fig:collinear}
as the propagators that become singular in the collinear limit 
in the virtual and real matrix elements are different. 
Reconsidering $p_i'$ as the parent parton momentum of the 
collinear splitting, we find the following relation 
between splitting matrices entering the real matrix elements
\beq
\sp^{(0) \dagger}(p_i', p_r'; \widetilde{p}_{ir}') = 
\frac{(\widetilde{p}_{ir}' - p_r')^2}{s_{ir}'} \, 
\sp^{(0)}(\widetilde{p}_{ir}', - p_r'; p_i')~, 
\label{antisplittings}
\eeq
where $(\widetilde{p}_{ir}' - p_r')^2/s_{ir}' = - n p_i'/n p_{ir}'$.
We show now that the factor $- n p_i'/n p_{ir}'$ is compensated 
by the phase-space. By introducing the following identity in 
the phase-space of the real corrections
\beq
1 = \int \, d^d p_{ir}' \, 
\delta^{(d)}\left(p_{ir}' - p_i' - p_r' \right)~,
\eeq
and performing the integration over the three-momentum ${\bf p}_i'$
and the energy component of $p_{ir}'$, the real 
phase-space becomes 
\beq
\int d\Phi_{N+1}(p_1; p_2', \ldots) = 
\int d\Phi_N(p_1; \ldots,p_{ir}',\ldots) \, 
\int_{p_r'} \, \td{p_r'} \, 
\frac{E_{ir}'}{E_i'}~,
\label{pscollinear}
\eeq
where the factor $(n p_i'/n p_{ir}')(E_{ir}'/E_i')$
equals unity in the collinear limit.
Inserting~\Eq{eq:piqicoll} in~\Eq{eq:I1ith}, 
and~\Eq{eq:piprimecoll}, \Eq{antisplittings} and \Eq{pscollinear} 
in \Eq{eq:I0ab} the loop and tree contributions show to have a 
very similar structure with opposite sign and match each other 
at the integrand level in the collinear limit. Correspondingly, 
soft singularities at $p_r'\to 0$ can be treated consistently as 
the endpoint limit of the collinear mapping.

\section{Conclusions and outlook}
\label{sec:conclusions}

The loop--tree duality method exhibits attractive theoretical
aspects and nice properties which are manifested by a direct physical 
interpretation of the singular behaviour of the loop integrand. 
Integrand singularities occurring in the intersection 
of on-shell forward hyperboloids or light-cones cancel among dual integrals. 
The remaining singularities, excluding UV divergences,
are found in the intersection of forward with backward on-shell 
hyperboloids or light-cones and are 
produced by dual propagators that are light-like or time-like separated 
and less energetic than the internal propagator that is set on-shell. 
Therefore, these singularities can be interpreted in terms of causality and 
are restricted to a finite region of the loop three-momentum space, 
which is of the size of the external momenta. As a result, 
a local mapping at the integrand level is possible between one-loop 
and tree-level matrix elements to cancel soft and collinear divergences. 
One can anticipate that a similar analysis at higher orders of the 
loop--tree duality relation is expected to provide equally interesting results. 
We leave this analysis for a future publication.

\section*{Acknowledgements}

We thank S. Catani for stimulating suggestions and 
the careful reading of the manuscript, 
and S. Jadach for very interesting discussions. 
This work has been supported by the 
Research Executive Agency (REA) of the European Union under 
the Grant Agreement number PITN-GA-2010-264564 (LHCPhenoNet),
by the Spanish Government and EU ERDF funds 
(grants FPA2011-23778 and CSD2007-00042 
Consolider Project CPAN) and by GV (PROMETEUII/2013/007). 
SB acknowledges support from JAEPre programme (CSIC).
GC acknowledges support from Marie Curie Actions (PIEF-GA-2011-298582).

\end{document}